\documentstyle[aps,prl,multicol,epsfig,array]{revtex}

\newcommand \be{\begin{eqnarray}}
\newcommand \ee{\end{eqnarray}}

\begin{document}

\title{Period doubling in glow discharges:
local versus global differential conductivity}

\author{Danijela D. \v{S}ija\v{c}i\'c$^1$, Ute Ebert$^{1,2}$ 
and Ismail Rafatov$^{1,3}$}

\address{$^1$CWI, P.O.Box 94079, 1090 GB Amsterdam, The Netherlands,\\
$^2$Dept. Physics, Eindhoven Univ. Techn., The Netherlands,\\
$^3$American University in Central Asia, Bishkek, Kyrgyzstan}

\date{\today}
\maketitle

\begin{abstract}
Short planar glow discharges coupled to a resistive layer exhibit 
a wealth of spontaneous spatio-temporal patterns. Several authors 
have suggested effective reaction-diffusion-models to explore
similarities with other pattern forming systems. To test these 
effective models, we here investigate the temporal oscillations 
of a glow discharge layer coupled to a linear resistor. 
We find an unexpected cascade of period doubling events. 
This shows that the inner structure of the discharge is more
complex than can be described by a reaction-diffusion-model
with negative differential conductivity. 
\end{abstract}

\begin{multicols}{2}

Glow discharges are part of our daily environment 
in conventional and energy saving lamps, beamers, flat TV screens, 
car and street lamps as well as in various industrial applications.
While applications typically try to avoid any instabilities, 
experiments actually exhibit a realm of spontaneous pattern formation,
see, e.g.,\cite{Kogel}.

An interesting series of experiments has been performed 
on short planar dc driven glow discharges with wide lateral extension 
\cite{Muenster0,Muenster1,Muenster2,StripeM,Muenster3,HexM,Zigstr,Str,filStr,Astr01,Str02,Muen03}
where the formed patterns were explored very systematically.
The observed structures resemble those observed in Rayleigh-Benard 
convection in flat cells, in electroconvection in nematic liquid 
crystals, or in various chemical or biological pattern forming systems. 
All these systems show the formation of stripes, spots, spirals etc.
In comparison to the other systems, the glow discharge system has 
the advantage of particular convenient experimental handling and 
time scales \cite{Gwinn}. Besides structures known from other
physical systems, it continues to exhibit new structures
that might be specific for this system
\cite{Muenster0,Muenster1,Muenster2,StripeM,Muenster3,HexM,Zigstr,Str,filStr,Astr01,Str02,Muen03}. 
We will focus on the experiment in \cite{Str},
where a complete phase diagram of different patterns
was identified: homogeneous stationary and
homogeneous oscillating modes, patterns with spatial and 
spatio-temporal structures etc.

The theory for these systems has largely focussed on effective 
reaction-diffusion models in the two transversal directions, 
and on the negative differential 
conductivity of the glow discharge as the driving force of
pattern formation. Such models actually have been developed
independently by a number of authors 
\cite{KGM,Rade89,Rade90,Rade92,Phelps93III,Petro97,Kolobov,Muenster2}.
On the other hand, the observation of unconventional patterns
like zigzag-destabilized spirals raises doubts whether
reaction-diffusion patterns are sufficient to understand 
the observed patterns.

In the present paper, we examine the concepts of reaction-diffusion
models and negative differential conductivity on the particular
case of a short DC driven glow discharge in a parameter range that exhibits 
spontaneous temporal oscillations but no spatial structures transverse 
to the current \cite{Str}.
In short, we find $~$ $(i)$ that a discharge on the transition from Townsend
to glow discharge can combine a positive local differential 
conductivity with a negative global differential conductivity;
$~$ $(ii)$ that a glow discharge in a simple electric circuit shows
more complex behavior than can be expected from the proposed
reaction-diffusion models 
\cite{KGM,Rade89,Rade90,Rade92,Phelps93III,Petro97,Kolobov,Muenster2}
for voltage $U$ and current $J$ with (global)
negative differential conductivity $dU/dJ<0$; 
$~$ $(iii)$ in particular, that the system can show period 
doubling bifurcations. 
Period doubling actually has been observed experimentally in
glow discharges, but in more complex geometries and in longer systems
\cite{PD1,PD2}.
$~$ $(iv)$ Finally, we derive a new effective dynamical model in terms 
of a parameter and a function by adiabatic elimination of the electrons. 
There is no systematic way to reduce this model to a simpler one
\cite{KGM,Rade89,Rade90,Rade92,Phelps93III,Petro97,Kolobov,Muenster2}
with two scalar parameters like voltage $U$ and current $J$.
We draw this conclusion both from direct analysis and from
the occurence of period doubling in the numerical solutions.

To be precise, in the experiments of 
\cite{Muenster0,Muenster1,Muenster2,StripeM,Muenster3,HexM,Zigstr,Str,filStr,Astr01,Str02,Muen03}, 
a planar glow discharge layer
with short length in the forward direction and wide lateral dimensions 
is coupled to a semiconductor layer with low conductivity. 
The whole structure is sandwiched between 
two planar electrodes to which a DC voltage $U_t$ is applied.
Theoretical predictions on how the different spatio-temporal
patterns depend on the parameters of the gas discharge, hardly exist.
In \cite{Rade89,Rade90,Rade92,Muenster2}, an effective 
reaction-diffusion model in the two dimensions
transversal to the current is proposed. Roughly, it consists of two
nonlinear partial differential equations for the current $J$ and 
the voltage $U$ of the form
\be
\label{RD}
\partial_tU(x,y,t)={\cal F}(U,J)~~~,~~~
\partial_tJ(x,y,t)={\cal G}(U,J)~,
\ee
where the nonlinear operators ${\cal F}$ and ${\cal G}$ contain spatial 
derivatives $\partial_x$, $\partial_y$ and possibly also integral kernels. 
The model is of
reactor-inhibitor form as studied extensively in the context of chemical 
and biological systems in the past decades. If applicable to gas discharges,
this identification lays a connection to a realm of analytical and 
numerical results on reaction-diffusion systems.

To test whether a model like (\ref{RD}) is applicable to the gas discharge
system, we will focus on its temporal oscillations 
that can occur in a spatially completely homogeneous mode \cite{Str}; 
hence a one-dimensional approximation is appropriate. Similar
oscillations have been observed in
\cite{Phelps93III,Petro97,Phelps93I,Pitch},
and similar effective models for current $J$ and voltage $U$ 
of the general form (\ref{RD}) have been proposed in 
\cite{KGM,Phelps93III,Petro97,Kolobov}.

Why have different authors come up with the same type of model?
The equation for $U$ directly results from the simplest form of
an external electric circuit: a semiconductor layer of thickness
$d_s$, linear conductivity $\sigma_s$ and dielectricity constant 
$\epsilon_s$ will evolve as
\be
\label{U_t}
\partial_tU=\frac{U_t-U-R_sJ}{T_s}
\ee
where $U_t$ is the voltage on the total system, $J$ is the total current, 
and $U=\int_0^{d_g}E\;dz$ is the voltage over the gas discharge which
is the electric field $E$ integrated in the $z$ direction over the 
height $d_g$ of the discharge. 
For the experiments in \cite{Str}, $R_s=d_s/\sigma_s$ is the resistance 
of the whole semiconductor layer, and $T_s=\epsilon_s\epsilon_0/\sigma_s
=C_sR_s$ is the Maxwell relaxation time of the semiconductor.
In other experimental systems, the quantities $R_s$ and $T_s$
can have different realizations. Hence the form of the equation 
for $U$ in a reaction-diffusion model (\ref{RD}) is clear.

However, the equation for $J$ in a reaction-diffusion model
as (\ref{RD}) is based on guesses and plausibility. Different
choices have been suggested by different authors, but one
thing is clear: to be physically meaningful, the current-voltage 
characteristics of the glow discharge has to be a stationary solution,
so ${\cal G}(U,J)=0$ on the characteristics. Beyond that,
there are different suggestions for the functional form 
of ${\cal G}$ and the intrinsic time scale.

If a model like (\ref{RD}) is applicable to oscillations
in glow discharge systems, then the following predictions apply:
$~$ 1) an oscillation can only occur in a region of negative 
differential conductivity of the glow discharge characteristics, $~$
2) only a single period can be formed, period doubling is not possible, 
since this requires at least three independent parameters, $~$
3) in a phase space plot in $U$ and $J$, the trajectory of an 
oscillation can intersect the load line $U=U_t-R_sJ$ only parallel
to the $J$-axis (since $\partial_tU=0$ and $\partial_tJ\ne0$), and 
it can intersect the characteristics of the glow discharge $U=U(J)$ 
only parallel to the $U$-axis (since $\partial_tU\ne0$ and $\partial_tJ=0$).

We now introduce the simplest classical model for a glow discharge
\cite{Engel,Raizer,DanaStat},
solve it numerically and confront its results with the predictions above.

A discharge between Townsend and glow regime consists of a gas
with Ohmic conductivity for the rare charged particles,
electrostatic space charge effects and two ionization mechanisms,
namely impact ionization by accelerated electrons in the bulk
of the discharge (the so-called $\alpha$-process) and secondary 
emission from the cathode (the $\gamma$-process).
In its simplest form, it can be modelled by continuity equations
for electron particle density $n_e$ and ion particle density $n_+$
\be
\label{1}
\partial_t\;n_e \;+\; \nabla\cdot{\bf J}_e
={\cal S} ~~~,~~~
\partial_t\;n_+ \;+\; \nabla\cdot{\bf J}_+
= {\cal S}~,
\ee
and the Poisson equation for the electric field {\bf E}
in electrostatic approximation, 
\be
\label{3}
\nabla\cdot{\bf E} = {{\rm e}\over{\varepsilon_0}} \;(n_+ -n_e)
~~~,~~~{\bf E}=-\nabla\Phi~. 
\ee
The particle currents are approximated as purely Ohmic
\be
\label{4}
{\bf J}_e = - \mu_e \;n_e \;{\bf E}~~~,~~~
{\bf J}_+ =  \mu_+\;n_+ \;{\bf E} ~. 
\ee
The source of particles in the continuity equation (\ref{1}) is written as 
a sum of generation by impact ionization in Townsend approximation
and recombination
\be
\label{5}
{\cal S} = |n_e \mu_e { E}| \;\alpha_0
\;\mbox{\large{e}}^{\textstyle -E_0/|{ E}|} - \beta n_e n_+ ~.
\ee
Finally, the secondary emission from the cathode enters as
a boundary condition at the position $d_g$ of the cathode
\be
\label{6}
\mu_en_e(d_g,t)=\gamma\mu_+n_+(d_g,t)~.
\ee
This is the classical glow discharge model \cite{Engel,Raizer,DanaStat}.

We reduce the problem to one spatial dimension $z$
transverse to the layers which is an excellent approximation for 
the experiment \cite{Str}. Furthermore, we introduce dimensionless 
quantities as in \cite{DanaStat} by rescaling all parameters 
and fields as $z=r_z/X_0$, $\tau=t/t_0$, $L=d_g/X_0$, 
$\sigma(z,\tau)=n_e(r_z,t)/n_0$, $\rho=n_+/n_0$, ${\cal E}=E_z/E_0$
with the scales $X_0=\alpha_0^{-1}$, $t_0=(\alpha_0\mu_eE_0)^{-1}$
and $n_0=\epsilon_0\alpha_0E_0/{\rm e}$. 
A key role is played by the small parameter $\mu=\mu_+/\mu_e$, 
which is the mobility ratio of ions and electrons.

The gas discharge layer is now modelled by
\be
\label{h01}
\partial_\tau\sigma&=&\partial_z({\cal E}\sigma)
+\sigma {\cal E}\alpha({\cal E})~~~,~~~\alpha({\cal E})=e^{-1/|{\cal E}|}~,\\
\label{h02}
\partial_\tau\rho&=&-\mu\partial_z({\cal E}\rho)
+\sigma {\cal E}\alpha({\cal E})~,\\
\label{h04}
&&\sigma(L,\tau)=\gamma\mu\rho(L,\tau)~,\\
\label{h05}
\rho-\sigma&=&\partial_z{\cal E}~,
\ee
where recombination was neglected [$\beta=0$ in (\ref{5}), a discussion of this
approximation follows below], while the external circuit is described by
\be
\label{h06}
\partial_\tau {\cal U}=
\frac{{\cal U}_t-{\cal U}-{\cal R}_s j}{\tau_s}~~,~~
\label{h07}
{\cal U}(\tau)=\int_0^L {\cal E}(z,\tau)\;dz
\ee
with the dimensionless voltage ${\cal U}=U/(E_0X_0)$, time scale 
$\tau_s=T_s/t_0$ and resistance ${\cal R}_s=R_s/R_0$, 
$R_0=X_0/({\rm e}\mu_en_0)$
and with a spatially conserved total current 
\be
\label{h08}
&&j(\tau)=\partial_\tau {\cal E}+\mu\rho {\cal E}+\sigma {\cal E}
~~,~~\partial_z j(\tau)=0~,
\ee
where $\partial_zj=0$ follows from 
(\ref{h01}), (\ref{h02}) and (\ref{h05}) as usual.

As a result, the gas discharge is parametrized by the three 
dimensionless parameters of system length over ionization length $L$, 
secondary emission coefficient $\gamma$ and mobility ratio $\mu$ 
(as discussed in \cite{DanaStat}), and the external circuit is
parametrized by relative resistance ${\cal R}_s$, ratio of time 
scales $\tau_s$ and dimensionless applied voltage ${\cal U}_t$.

For calculational purposes, the ion density $\rho$ can be
completely eliminated from the one-dimensional gas discharge equations 
(\ref{h01})--(\ref{h05}) with the help of the Poisson equation (\ref{h05})
and the total current $j$, see \cite{DanaStat}. The result are two equations
of motion for $\partial_\tau\sigma$ and $\partial_\tau{\cal E}$.
In our numerical calculations, the system was implemented in
this form. Our choice of parameters was guided by
the experiments in \cite{Str}: we chose the secondary emission coefficient
$\gamma=0.08$, the mobility ratio $\mu=0.0035$ for nitrogen and
the dimensionless system size $L=50$ which amounts to 1.4 mm 
at a pressure of 40 mbar. The external circuit has ${\cal R}_s=30597$, 
$\tau_s=7435$ and a dimensionless total
voltage ${\cal U}_t$ in the range between 18 and 20.
This corresponds to a GaAs layer with $\epsilon_s=13.1$, conductivity
$\sigma_s=(2.6\cdot 10^5\Omega{\rm cm})^{-1}$ and thickness 
$d_s=1.5 {\rm mm}$, and a voltage range between 513 and 570 V.

\begin{figure}[h]
\centerline{
\psfig{figure=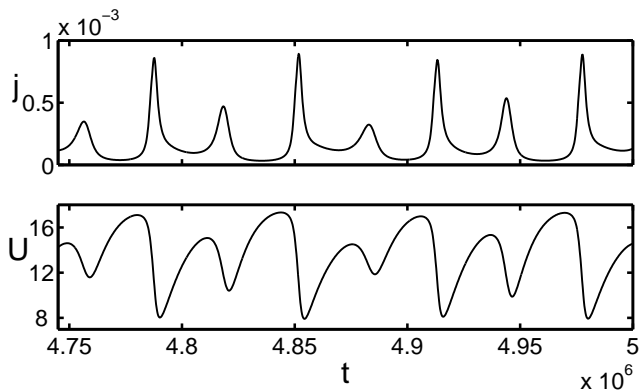,width=8.5cm}}
\caption{Spontaneous oscillations of current $j$ and voltage ${\cal U}$ 
as a function of time $\tau$ for
$\gamma=0.08$, $\mu=0.0035$, $L=50$, ${\cal R}_s=30597$, $\tau_s=7435$, 
and applied total voltage ${\cal U}_t=19$.}
\end{figure}

Fig.~1 shows electric current $j$ and voltage on the gas discharge
${\cal U}$ as a function of time for a total stationary voltage
${\cal U}_t=19$ applied to the complete system of gas discharge and
semiconductor layer. The system exhibits spontaneous oscillations 
with sharp current peaks: when the voltage ${\cal U}$ on the gas layer
becomes high enough, the discharge ignites. The conductivity of 
the gas increases rapidly and produces a current pulse that deposits 
a surface charge on the gas-semiconductor interface. Therefore the voltage 
${\cal U}$ over the gas layer breaks down. Due to the low conductivity of the
semiconductor, the voltage ${\cal U}$ recovers only slowly. Eventually
the gas dicharge ignites again, and the cycle is repeated.

Note that the oscillations in Fig.~1 are not quite periodic. 
This is not due initial transients since the system is observed
after the long relaxation time $\tau=4.745\cdot10^6$. The nature
of this temporal structure becomes clear when the trajectory
is plotted in the plane spanned by current $j$ and voltage ${\cal U}$
in Fig.~2(b). The figure contains the data of the time span
from $\tau=3\cdot10^6$ to $6\cdot10^6$ which amounts to approximately 90 
current pulses.
The phase space plot shows that the system is actually periodic,
with a period of 8 current pulses. Fig.~1 shows precisely one period.

This discovery raises the question whether our system actually
follows the well-known scenario of period doubling. Indeed, it does. 
Fig.~2(a) for ${\cal U}_t=18$ shows an oscillation where one current pulse
is repeated periodically as observed experimentally in \cite{Str}.
For ${\cal U}_t=18.5$, a period consists of two current pulses (not shown).
For ${\cal U}_t=19$, the period is 8 pulses as in Fig.~1 and Fig.~2(b).
For ${\cal U}_t=20$, the systems seems to have reached the chaotic state
as can be seen in Fig.~2(c).

A detailed comparison of the experiments in \cite{Str} with simple 
oscillations as in Fig.~2(a) will be given elsewhere, and we only state 
here that there is semi-quantitative agreement of several features.
Here we emphasize that period doubling events 
in glow discharges have been observed experimentally in other systems
\cite{PD1,PD2}. However, this was always in systems with
more complicated geometries like long narrow tubes, and the authors allude
to general knowledge on nonlinear dynamics rather than to solutions of
explicit models. We state that period doubling can be a generic 
feature of a simple, strictly one-dimensional glow discharge 
when coupled to the simple circuit (\ref{U_t}). We propose to search 
experimentally for a period doubling route to chaos in such simple 
systems which would then allow quantitative comparison with theory.

Let us return to
the initial question: is a 2-component reaction diffusion model
like (\ref{RD}) with negative differential conductivity appropriate 
for the present system? Above Eq.~(\ref{1}), we gave a list of predictions
for the reaction diffusion model (\ref{RD}) to be applicable. 
Prediction 2 is falsified by the observation of period doubling.
Prediction 3 is also falsified by a simple check of either of the 
three figures in Fig.~2: the trajectories definitely do not intersect
with the characteristics or the load line with the angle prescribed 
by (\ref{RD}), in particular not in the upper part of the figures
that represent the rapid current pulses.

There rests prediction 1: is negative differential conductivity required
for the oscillations to occur? We have not found a numerical 
counterexample where oscillations would occur while the current voltage
characteristics of the gas discharge shows a positive differential 
conductivity, but we have found no reason to exclude its existence. 
Furthermore we note that the characteristics is a global property of 
the whole discharge layer with its boundary conditions \cite{DanaStat}
while the local differential conductivity in our model is always
positive: the field dependent stationary ionization is 
$n_+=|\mu_eE|\alpha_0 e^{-E_0/|E|}/\beta$ according to (\ref{5}); hence 
the local conductivity increases monotonically with the applied field $|E|$.
The global negative differential conductivity is due to electrode
effects being much stronger than bulk recombination $\beta$.

Last but not least, we have derived an analytical 
approximation of the model (\ref{h01})--(\ref{h08}) that can be 
confronted with the suggested form (\ref{RD}). Electron and ion current
in the gas are of the same order of magnitude. Since the electrons 
are much more mobile, their density is appropriately lower.
Rescaling this density like $s=\sigma/\mu$ and time like
$\bar\tau=\mu\tau$, the electrons can be eliminated adiabatically
in the limit of $\mu\to0$. Space charges in the gas discharge are 
then due to the ions only $\rho=\partial_z{\cal E}$, and $\rho$ 
can be expressed by ${\cal E}$. Splitting the field 
${\cal E}(z,t)={\cal E}_L(t)+\epsilon(z,t)$ into the field on the cathode 
${\cal E}_L$ and a correction $\epsilon$ with $\epsilon(L,t)=0$,
the complete system for $\mu\to0$ can be expressed by two dynamic equations
\be
\label{RDus}
\partial_{\bar\tau}{\cal E}_L(t)=F({\cal E}_L,\epsilon)~~~,~~~
\partial_{\bar\tau}\epsilon(z,t)=G({\cal E}_L,\epsilon)~,
\ee
details will be given elsewhere.
While the equation for the time dependent parameter ${\cal E}_L$
corresponds to the equation for $U$ in (\ref{RD}), the space dependent
field correction $\epsilon$ within the gas layer cannot be reduced to
a single component like the current $J$ in (\ref{RD}). E.g., for the ion
density on the cathode $\rho_L=\partial_z\epsilon|_L$, 
we can derive the equation of motion
\be
\label{adia}
\partial_{\bar\tau}\rho_L
=-\rho_L^2-{\cal E}_L(\partial_z\rho)|_L
+\gamma\rho_L{\cal E}_L\alpha({\cal E}_L)
\ee
A two component reaction-diffusion equation for $\rho_L$ and ${\cal E}_L$
could result from the completely unsystematic approximation 
$(\partial_z\rho)|_L=0$. Rather the transport of ions $\rho$ 
from the bulk of the gas towards the cathode 
is a central feature of the system. The field 
$\epsilon(z,t)$ in (\ref{RDus}) indeed accounts for 
the ion distribution within the gas gap with its intricate dynamics.

We acknowledge support of D.S. by the Dutch physics
funding agency FOM and of I.R. by ERCIM.

\end{multicols}

\begin{figure}[h]
\centerline{
\psfig{figure=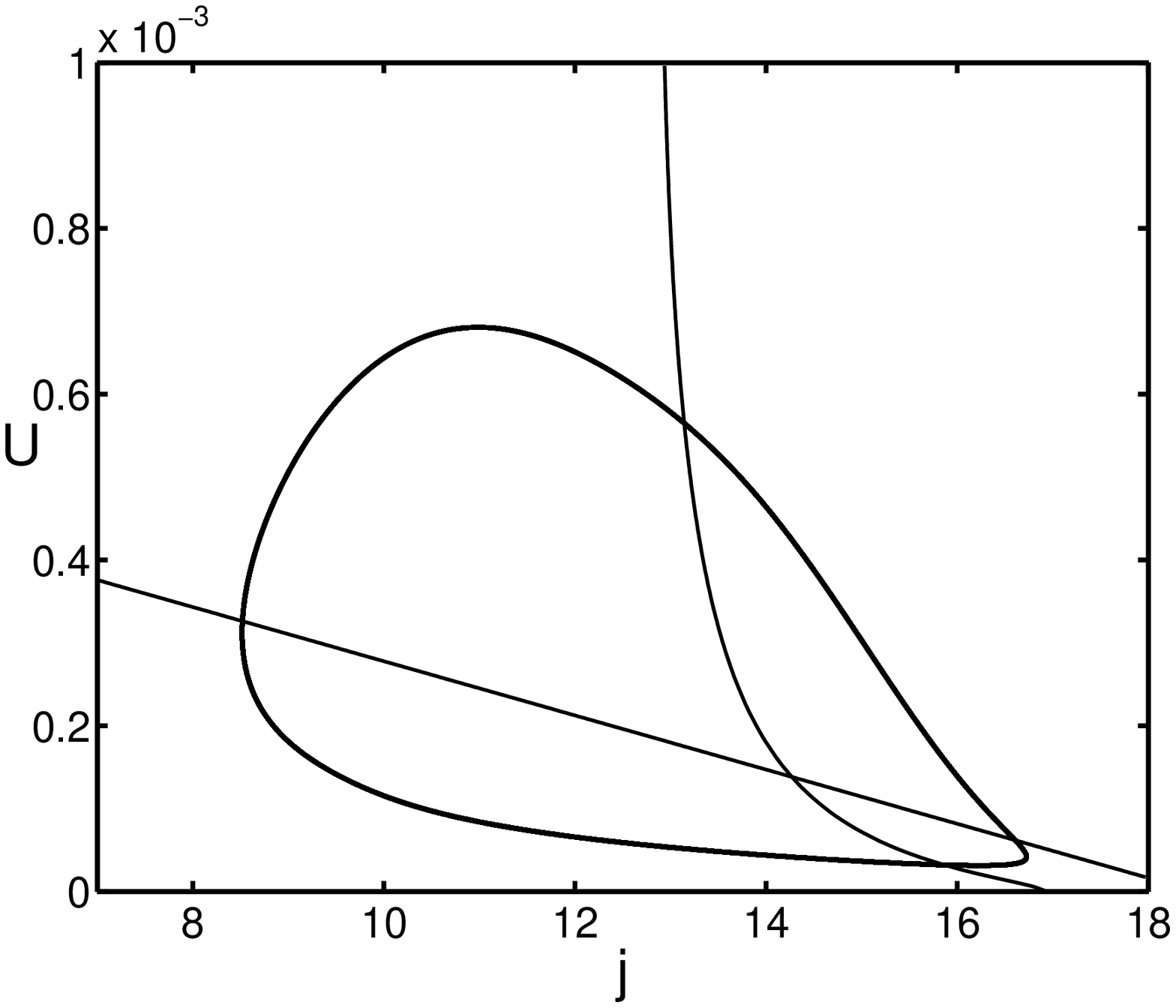,width=6cm}
\psfig{figure=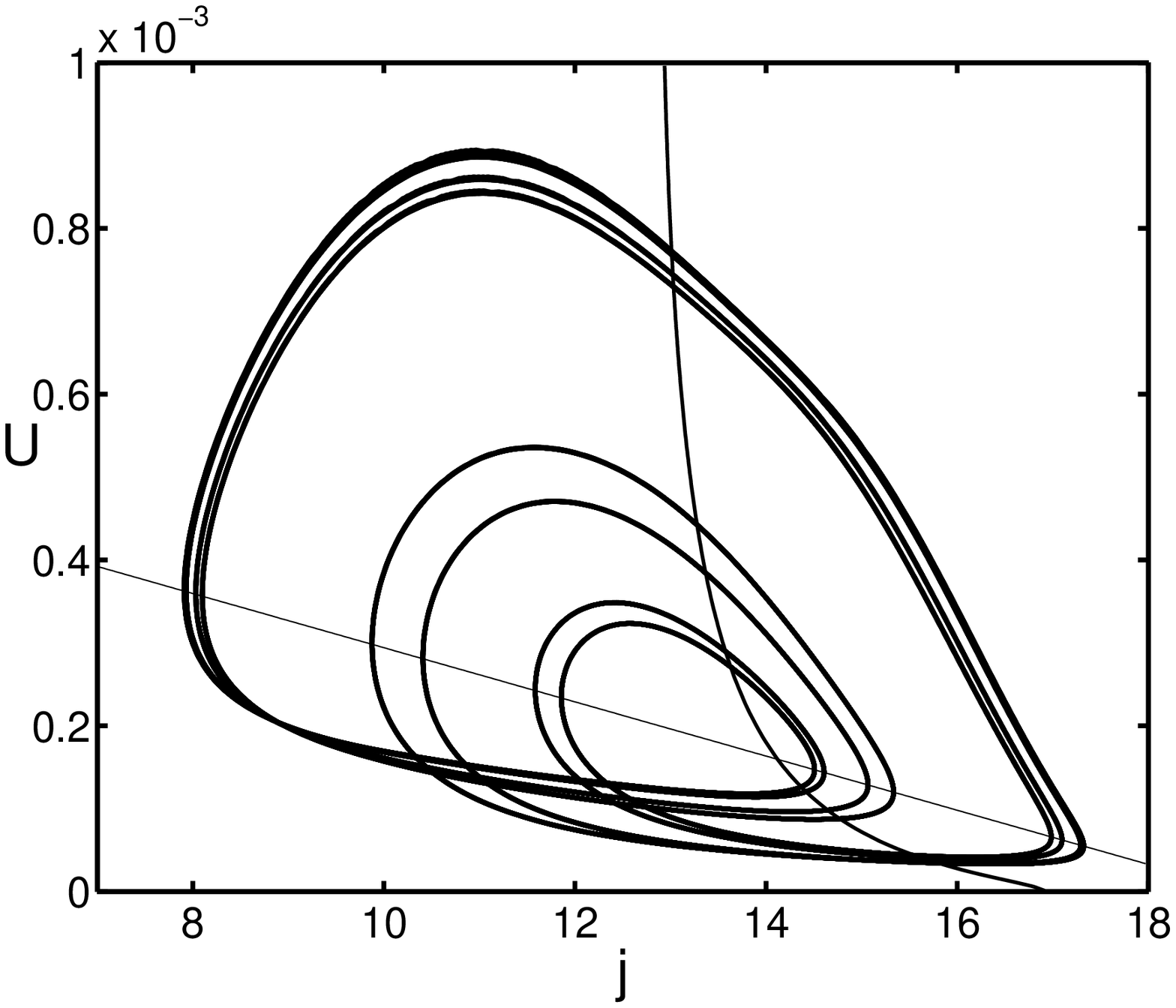,width=6cm}
\psfig{figure=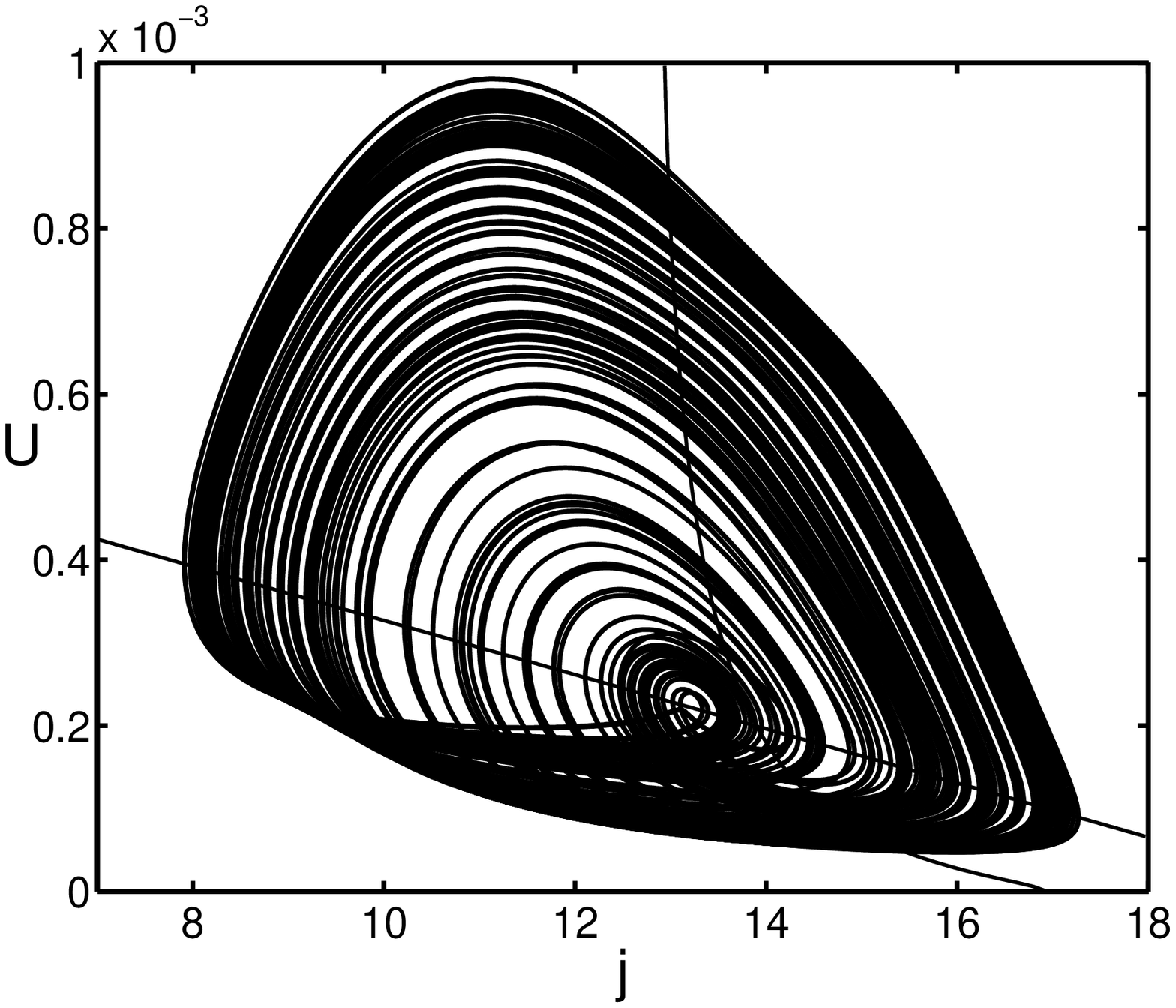,width=6cm}
            }
\caption{Phase space plots of the trajectories of the oscillations 
in the plane of current $j$ and voltage ${\cal U}$.
The time range is $3\cdot10^6\le\tau\le6\cdot10^6$ in all figures.
Shown are the orbits, the straight load line ${\cal U}={\cal U}_t-{\cal R}_sj$
and the curved current voltage characteristics ${\cal U}={\cal U}(j)$ of 
the gas discharge [28]. 
The intersection of load line and characteristics 
marks the stationary solution of the system.
(b) represents the data of Fig.~1 with total voltage ${\cal U}_t=19$,
(a) is for ${\cal U}_t=18$, (c) for ${\cal U}_t=20$.}
\end{figure}

\end{document}